\documentclass[aps, prd, preprintnumbers,nofootinbib,notitlepage]{revtex4-1}

\usepackage{colortbl}
\usepackage[T1]{fontenc}
\usepackage{microtype}
\usepackage[textsize=scriptsize,backgroundcolor=red!70,linecolor=red]{todonotes}
\usepackage{booktabs}
\usepackage{dcolumn}
\usepackage{amsmath}
\usepackage{amsfonts}
\usepackage{amssymb}
\usepackage{graphicx}
\usepackage{hyperref}
\usepackage[all]{hypcap}
\usepackage{paralist}
\usepackage{multirow}

\usepackage{graphicx}
\usepackage{dcolumn}
\usepackage{bm}
\usepackage{epsf}
\usepackage{dcolumn}
\def\be{\begin{equation}}
\def\ee{\end{equation}}
\def\bea{\begin{eqnarray}}
\def\eea{\end{eqnarray}}\def\nn{\nonumber}
\def\gsim{\ \rlap{\raise 2pt\hbox{$>$}}{\lower 2pt \hbox{$\sim$}}\ }
\def\lsim{\ \rlap{\raise 2pt\hbox{$<$}}{\lower 2pt \hbox{$\sim$}}\ }
\def\dslash{\kern-4pt \not{\hbox{\kern-2pt $\partial$}}}
\def\pslash{\not{\hbox{\kern-2pt p}}}

\def\l{{\rm L}}

\definecolor{gray}{rgb}{0.90,1,1}
\definecolor{LightCyan}{rgb}{0.88,1,1}

\def\bsmumu{b \to s \mu^+ \mu^-}

\def\bsee{b \to s e^+ e^-}
\def\bsll{b \to s \ell^+ \ell^-}

\def \SM{{\rm SM}}

\def \expt{{\rm expt}}

\def \be{\beta}

\def\bs{B^0_s}
\def\bsbar{{\bar B}^0_s}

\def\beq{\begin{equation}}
\def\eeq{\end{equation}}
\def\bea{\begin{eqnarray}}
\def\eea{\end{eqnarray}}
\def\ber{\begin{eqnarray*}}
\def\eer{\end{eqnarray*}}
\def\bwt{\begin{widetext}}
\def\ewt{\end{widetext}}
\def\nn{\nonumber}
\def\roughly#1{\mathrel{\raise.3ex\hbox
{$#1$\kern-.75em\lower1ex\hbox{$\sim$}}}}
\def\lsim{\roughly<}
\def\gsim{\roughly>}

\def\order{\lower 1.8ex \hbox{\LARGE\~{}}}

\def\RK{R_K}

\def\RKstar{R_{K^*}}

\usepackage{bm}

\newcommand{\bctaunutau}{b \to c \tau^- {\bar\nu}_\tau}

\def \({\left(}
\def \){\right)}
\def \[{\left[}
\def \]{\right]}
\def \l|{\left|}
\def \r|{\right|}

\def \nn{\nonumber}

\def \be{\beta}

\def \SM{{\rm SM}}

\def \expt{{\rm expt}}

\def\Z{Z^{\prime}}

\def\bsmumu{ b \to  s \mu^+ \mu^-}

\def\bs{B_s^0}

\def\bsbar{{\bar B}_s^0}

\def\RK{R_K}

\def\RKstar{R_{K^*}}

\begin{document}
\DeclareGraphicsExtensions{.eps,.ps}


\title{New light mediators for the $R_K$ and $R_{K^*}$ puzzles}


\author{Alakabha Datta}
\affiliation{Department of Physics and Astronomy, University of Mississippi, 108 Lewis Hall, Oxford, MS 38677, USA}
\affiliation{Department of Physics and Astronomy, University of Hawaii at Manoa, Honolulu, HI 96822, USA}

\author{Jacky Kumar}
\affiliation{Department of High Energy Physics, Tata Institute of Fundamental Research, Mumbai 400005, India}

\author{Jiajun Liao}
\affiliation{Department of Physics and Astronomy, University of Hawaii at Manoa, Honolulu, HI 96822, USA}
 
\author{Danny Marfatia}
\affiliation{Department of Physics and Astronomy, University of Hawaii at Manoa, Honolulu, HI 96822, USA}

\begin{abstract}
The measurements of $R_K$ and $R_{K^*}$ provide hints for the violation of lepton universality. 
However, it is generally difficult to explain the $R_{K^*}$ measurement in the low $q^2$ range, 
$0.045 \le q^2 \le 1.1~{\rm GeV}^2$. 
Light mediators offer a solution by making the Wilson coefficients $q^2$ dependent.
We check if new lepton nonuniversal interactions mediated 
by a scalar ($S$) or vector particle ($\Z$) of mass between $10-200$~MeV can reproduce the data.
We find that a 25 MeV $\Z$ with a $q^2$-dependent $b-s$ coupling and that couples to the electron but not the muon can explain all three anomalies 
 in conjunction with other measurements. A similar 25 MeV $S$ provides a good fit to all relevant data except $R_{K^*}$ in the low $q^2$ bin. A 25 MeV $\Z$ with a $q^2$-dependent $b-s$ coupling and that couples to the muon but not the electron provides a good fit to the combination of the $R_K$ and $R_{K^*}$ data, but does not fit $R_{K^*}$ in the low $q^2$ bin well.
  
\end{abstract}
\pacs{14.60.Pq,14.60.Lm,13.15.+g}
\maketitle

\section{Introduction}
The search for new physics in $B$ decays is an ongoing endeavor. Recently,
anomalies in semileptonic $B$ decays have received a lot of attention. These anomalies are found in the charged current $ \bctaunutau$ and neutral current $\bsll$ transitions.
Here we focus on the neutral current anomalies though the anomalies might be related~\cite{datta_shiv}. 
 Other anomalies appear in
 $B \to K^* \mu^+\mu^-$ where the LHCb~\cite{BK*mumuLHCb1,BK*mumuLHCb2} and 
 Belle~\cite{BK*mumuBelle}
  Collaborations find  deviations from the Standard model (SM) predictions, particularly
  in the angular observable $P'_5$ \cite{P'5}.
 The ATLAS~\cite{BK*mumuATLAS} and CMS \cite{BK*mumuCMS} Collaborations
  have also made measurements of the $B \to K^*
  \mu^+\mu^-$ angular distribution with results consistent with LHCb. Further, the
  LHCb has made measurements of the branching ratios and angular distributions in
   $\bs \to \phi \mu^+ \mu^-$~\cite{BsphimumuLHCb1,BsphimumuLHCb2} which are at variance with SM predictions
   based on lattice QCD~\cite{latticeQCD1,latticeQCD2} and QCD sum rules
  \cite{QCDsumrules}.
 
 The measurements discussed above are subject to unknown hadronic uncertainties \cite{silves} making it necessary to construct clean observables to test for new physics (NP).
One such observable is $R_K \equiv {\cal B}(B^+ \to K^+ \mu^+ \mu^-)/{\cal
  B}(B^+ \to K^+ e^+ e^-)$~\cite{hiller1, hiller2}, which has been measured by
 LHCb~\cite{RKexpt}:
\beq
R_K^\expt = 0.745^{+0.090}_{-0.074}~{\rm (stat)} \pm 0.036~{\rm (syst)} ~,~~ 1 \le q^2 \le 6.0 ~{\rm GeV}^2\,.
\label{RKexpt}
\eeq
This differs from the SM prediction, $R_K^\SM = 1 \pm 0.01$~\cite{IsidoriRK} by $2.6\sigma$.
Note, the observable $R_K$ is a measure of lepton flavor universality and requires different new physics for the muons versus the electrons, while it is possible to explain the anomalies in the angular observables in $\bsmumu$ in terms of lepton flavor universal new physics~\cite{Datta:2013kja}.

Recently, the LHCb Collaboration reported the measurement of the
ratio $R_{K^*} \equiv {\cal B}(B^0 \to K^{*0} \mu^+ \mu^-)/{\cal
  B}(B^0 \to K^{*0} e^+ e^-)$ in two different ranges of the dilepton
invariant mass-squared $q^2$~\cite{RK*expt}:
\beq
R_{K^*}^\expt = 
\left\{ 
\begin{array}{cc}
0.660^{+0.110}_{-0.070}~{\rm (stat)} \pm 0.024~{\rm (syst)} ~,~~ & 0.045 \le q^2 \le 1.1 ~{\rm GeV}^2 ~, \ \ \ (\rm{low} \ q^2)\\
0.685^{+0.113}_{-0.069}~{\rm (stat)} \pm 0.047~{\rm (syst)} ~,~~ & 1.1 \le q^2 \le 6.0 ~{\rm GeV}^2 ~, \ \ \ (\rm{central}\ q^2)\,.
\end{array}
\right.
\eeq
These differ from the SM predictions
by 2.2-2.4$\sigma$ (low $q^2$) and 2.4-2.5$\sigma$
(central $q^2$), which  further strengthens the hint  of lepton
non-universality observed in $R_K$. 

Lepton universality violating new physics may occur in $\bsmumu$ and/or
$\bsee$ transitions. 
The fact that the measurement of ${\cal
  B}(B^+ \to K^+ e^+ e^-)$ is found to be consistent with the
prediction of the SM may lead one to conclude that
 NP is more likely to be in
$\bsmumu$. However,  the branching ratios suffer from hadronic uncertainties~\cite{bslltheorerror} unlike the ratios $R_K$ and $R_{K^*}$
and so  new physics in
$\bsmumu$  and/or in $\bsee$ is still allowed.

Since the announcement of the $R_{K^*}$ result, a number of papers have analyzed the new measurements, mostly in terms of new physics with heavy mediators~\cite{Capdevila:2017bsm,Altmannshofer:2017yso,DAmico:2017mtc,Hiller:2017bzc,Geng:2017svp,Ciuchini:2017mik,Celis:2017doq,Becirevic:2017jtw,DiChiara:2017cjq, Sala:2017ihs,Ghosh:2017ber,datta_rk,jacky_rk,wang, Bonilla:2017lsq}.
The general conclusion is that there is a significant disagreement with
the SM, possibly as large as $\sim 6\sigma$,  and 
that theoretical hadronic uncertainties~\cite{BK*mumuhadunc1,BK*mumuhadunc2,BK*mumuhadunc3}  are insufficient to understand the data. 
However, with heavy new physics it is difficult to understand the $R_{K^*}$ measurement  in the very low $q^2$ bin
$0.045 \le q^2 \le 1.1 ~{\rm GeV}^2$, although the predictions are consistent with measurements within  $ 1.5  \sigma$. A  resolution to
this problem may be possible if the new physics is light.

In models with light mediators~\cite{ZMeV,Sala:2017ihs, Ghosh:2017ber, datta_rk, Bishara:2017pje}, the new physics
 cannot be integrated out, resulting in a $q^2$ dependence of
the Wilson coefficients (WCs).
If the light mediator mass is between $m_B$ and twice the lepton mass, 
and the mediator width is narrow, then it is observable 
as a resonance in the dilepton invariant mass. 
To avoid constraints from the search for such states, one generally 
takes the mediator mass to be $m_B$ or less than
 $2m_\ell$. In this paper we study a light scalar mediator denoted by $S$ and a 
light vector mediator denoted by $\Z$.
 
\section{Light scalar}
We start our discussion with a light scalar $S$ with mass in the $10-200$ MeV range.
For this scenario, we assume the following flavor-changing $bsS$ vertex, 

\beq
F(q^2) \,\bar{s} \left[ g_{bs}^{S} P_L + g_{bs}^{S'} P_R \right]  b \, S  ~,
\label{bsS}
\eeq
where $F(q^2)$ is a form factor.{\footnote{In our effective theory approach, the structure in Eq.~(\ref{bsS}) is of the general form consistent with the assumed symmetries. As an illustration of how a flavor changing vertex with a $q^2$-dependent form factor may occur, consider the following Lagrangian at the $b$-quark mass scale in the gauge basis:
\bea
\cal{L} & =&\frac{g}{\Lambda^2} \bar{b} b \bar{\chi} \chi + g_\chi \bar{\chi} \chi S\,,\
\eea
where $\chi$ is a hidden sector fermion (which may serve as a dark matter candidate) of mass $m_{\chi} \lsim m_b$, 
and we have suppressed all Lorentz structures in the Lagrangian. 
(In the context of Section~III, for a light vector mediator $\Z$, one may consider a similar Lagrangian of the form,
$g_\chi \bar{\chi} \gamma^{\mu} \chi Z^{\prime}_{\mu}$.)
The first term in the Lagrangian represents an effective coupling between the $b$ and $\chi$ fields that 
might arise via the exchange of a heavy mediator of mass $\Lambda \gg m_b$, which has been integrated out of the theory at the $m_b$ scale. Although there is no direct coupling between $b$ and $S$ (or $\Z$),  a $\bar{b}b S$ (or
 $\bar{b}b \Z$)  vertex with a $q^2$-dependent coupling will be generated by a $\chi$ loop. Transforming the $b$ quark from the  gauge to the mass basis then generates a $ \bar{s} b S$ (or $\bar{s}b \Z$) coupling.
In the case of the scalar mediator the form factor contains terms  of the form,
$ \frac{m_{\chi}^2}{\Lambda^2}$ and  $ \frac{q^2}{\Lambda^2}$. For the latter term to dominate, $q^2\gg m_\chi^2$, which implies that $m_\chi\lsim 30$~MeV for the $q^2$ values of interest.
For the $\Z$ case, the leading term in the form factor goes as $q^2$ due to the conserved vector current~\cite{FF}. 
We note that
 the situation is similar to the SM case where $\chi$ is replaced by the charm quark and $S$ (or $\Z$) by the photon. In this case  the first term in the Lagrangian, of the form
$\frac{g}{M_W^2} \bar{b} s \bar{c} c $,
is just one of the terms in the SM effective Lagrangian after integrating out the $W$ boson. The charm loop then induces an effective $ \bar{b} s \gamma^*$  vertex which yields   $ \bar{b} s \ell^+ \ell^-$ via $\gamma^* \to \ell^+ \ell^-$.}}
The matrix  elements for the  processes $\bsll$ and the mass difference in $B_s$ mixing are
\bea
M_{\bsll} 
& = & 
\frac{ F(q^2)}{q^2- M_{S}^2} \left [\bar s ( g_{bs}^{S}P_L + g_{bs}^{S'} P_R ) b \right]
\left [ \bar \ell (g_{L}^{\ell \ell}P_L + g_{R}^{\ell \ell}P_R) \ell \right ] ~, \nn\\
\Delta M_{{B_s}}^{NP} & = &
\frac{(F(q^2))^2}
  {2 q^2- 2 M_{S}^2}    f_B^2{m_{B_s}}
    \left[ -\frac{5}{12}\left( {(g_{bs}^{S})}^2 +{(g_{bs}^{S'})}^2\right) 
    +2 g_{bs}^Sg_{bs}^{S'} \frac{7}{12}\right], \
\label{meS}
\eea
where we have used Ref.~\cite{soni_2HDM} for $\bs$-$\bsbar$ mixing. 
The mass difference  in the SM for the $B_s$ system is
\cite{newpaper_datta}
\beq
\Delta M_{B_s}^{SM} = (17.4 \pm 2.6)~{\rm ps}^{-1}\,,
\eeq
which is consistent with experimental measurement~\cite{HFAG},
\beq
\Delta M_{B_s} = (17.757 \pm 0.021)~{\rm ps}^{-1}\,.
\eeq
We will choose the new physics contribution, 
$\Delta M_{B_s}^{NP}$, to be as large as the uncertainty in the SM  prediction.
 
We now consider $\bsll$ transitions. 
For light scalars coupling to muons, $\RK$ and $\RKstar$ are generally increased from their SM values in contradiction with experiment. Moreover, the measured $B_s \to \mu^+ \mu^-$ rate also puts strong constraints on new
scalar couplings to muons.

 We therefore suppose the scalar couples mainly to electrons 
 in which case the matrix element for $\bsee$  from Eq.~(\ref{meS}) is 
\bea
M_{\bsee}^{S,S'}  = 
 \frac{ g_{ee}^S}
{q^2- M_S^2} F(q^2)\left[ g_{bs}^S(\bar{s} P_L b)  + {g_{bs}^{S'} }
   (\bar{s} P_R b)\right]  (\bar{e}  e) +
  \frac{ g_{ee}^{S'}}
{q^2- M_S^2} F(q^2)\left[ g_{bs}^S(\bar{s} P_L b)  + {g_{bs}^{S'} }
   (\bar{s} P_R b)\right]  (\bar{e} \gamma_5 e)  
    \,,\
\eea
where $g_{ee}^S\equiv(g_{L}^{ee} + g_{R}^{ee})/2$ and
 $g_{ee}^{S'}\equiv(g_{R}^{ee} -g_{L}^{ee})/2$. In the following discussion, we chose different structures for the form factor $F(q^2)$.
 
\subsection{$F(q^2)\equiv 1$}
First, we consider the situation in which the $bsS$ vertex is generated either at tree level or at loop level with internal particles with masses much greater than the $b$ quark mass.
Then, the form factor $F(q^2)\equiv 1$, and to avoid a pole contribution to the measurements
  of ${B(B^0\to K^{*0}e^+e^-)}$
 in the dielectron invariant mass range,
  $ m_{ee}=[30 -1000 ]$~MeV~\cite{Aaij:2013hha}, we choose $M_S= 25$~MeV. 
 
  Note that the BaBar~\cite{Aubert:2008ps} and Belle~\cite{adachi, wei} measurements require $m_{ee}$ to be larger than 30~MeV~\cite{flood} and 
140~MeV, respectively. 
We fix $g_{ee}^S=2.0\times 10^{-4}$, which is the largest value allowed by the anomalous magnetic moment of the
electron~\cite{ae} for $M_S= 25$ MeV at the 2$\sigma$ CL.
Then we perform a $\chi^2$-fit to the 
theoretically clean observables $R_K$ and $R_{K^*}$, and the new physics contribution to the $B_s$ mass difference,  $\Delta M_s^{NP}=0\pm 2.6$~ps$^{-1}$. 
In Ref.~\cite{Wehle:2016yoi} the lepton flavor dependent 
angular observables $Q_{4,5}$ were measured but since the errors 
in the measurements are large we do not use them in our 
fit. 
We use {\tt flavio}~\cite{flavio} to calculate 
the theoretical values of the observables 
$\mathcal{O}_{th}$. We then compute
\beq
\chi^2(g_{bs}^S,g_{bs}^{S'}) = \sum_{R_K, R_{K*},\Delta M_s^{NP}} (\mathcal{O}_{th}(g_{bs}^S,g_{bs}^{S'}) -\mathcal{O}_{exp})^T \, \mathcal{C}^{-1} \,
(\mathcal{O}_{th}(g_{bs}^S,g_{bs}^{S'}) -\mathcal{O}_{exp})\,,
\eeq
where $\mathcal{O}_{exp}$ are the experimental measurements of the
observables, and the total covariance matrix $\mathcal{C}$ is 
the sum of theoretical and experimental covariance matrices. 
The SM gives a very poor fit to the $R_K$ and $R_{K^*}$ measurements with
\begin{equation} \chi^2_\text{SM} / \text{dof}=25.5/3\,. \end{equation}

\begin{table}[t]
\label{tab:Zprime}
\label{tab:scalar}
\fontsize{10}{10}\selectfont{
\caption{The fit results and the predictions for $R_K$ and $R_{K^*}$ at the best fit point for three scenarios of a light mediator with a mass of $25$~MeV. }
\begin{center}
\resizebox{\columnwidth}{!}{%
\renewcommand{\arraystretch}{1.5}
\begin{tabular}{|c|c|c|c|c|c|c|}\hline

Case &  &  & ${R_{K^*}}_{[0.045-1.1]}$ &  ${R_{K^*}}_{[1.1-6.0]}$   &  ${R_K}_{[1.0-6.0]}$  & pull \\ \hline
\multicolumn{3}{|c|}{Experimental results   } & $0.66 \pm 0.09$ &  $0.69 \pm 0.10$   &  $0.75 \pm 0.09$ & \\ \hline
\multicolumn{3}{|c|}{Standard model predictions   } & 0.93 &  0.99   &  1.0 &  \\ \hline
\multicolumn{7}{|c|}{ (i) Light scalar with electron coupling  }   \\ \hline
$F(q^2) \equiv 1$, $g_{ee}^S=2.0\times 10^{-4}$ & $g_{bs}^S g_{ee}^S= (12.6 \pm 2.2 ) \times 10^{-9}$ & $g_{bs}^{S'}g_{ee}^S = (4.0 \pm 1.6 ) \times 10^{-9}$ & 0.70 & 0.91 & 0.69 & 4.3 \\ \hline
$a_{bs} \neq 0$ & $g_{bs}^S g_{ee}^S = (-1.3 \pm 2.1 ) \times 10^{-9}$ & $g_{bs}^{S'}g_{ee}^S = (-13.1 \pm 2.1 ) \times 10^{-9}$ & 0.58 & 0.85 & 0.75 & 4.7 \\ \hline
$a_{bs} = 0$ & $g_{bs}^S g_{ee}^S = (2.7 \pm 2.6 ) \times 10^{-8}$ & $g_{bs}^{S'}g_{ee}^S = (-15.5 \pm 2.6 ) \times 10^{-8}$ & 0.89 & 0.65 & 0.75 & 4.4 \\ \hline
\multicolumn{7}{|c|}{ (ii) Light vector with muon coupling  }   \\ \hline
$F(q^2) \equiv 1$, $ g_{L}^{\mu\mu} =  g_{R}^{\mu\mu}=8.0\times 10^{-4}$ & $g_{bs}g_{\mu\mu} = (2.3 \pm 2.0) \times 10^{-10}$ & $g_{bs}'g_{\mu\mu} = (1.3 \pm 2.2 ) \times 10^{-10}$ & 0.93                        &0.99                        &0.96  & 1.4\\ \hline
$a_{bs} \neq 0$, $ g_{L}^{\mu \mu} =  g_{R}^{\mu \mu}$  & $ g_{bs}g_{\mu \mu} = (6.5 \pm 3.5) \times 10^{-10}$ & $ g_{bs}'g_{\mu \mu} = (1.6 \pm 3.6 ) \times 10^{-10}$ & 0.93                        &0.96                        &0.92  & 2.4\\ \hline
$a_{bs} \neq 0$, $g'_{bs}=0$, $ g_{L}^{\mu \mu} \neq  g_{R}^{\mu \mu}$  & $ g_{bs} g_{\mu\mu}=(5.7\pm 2.3) \times 10^{-10}$ & $ g_{bs}g'_{\mu\mu}=(0.2 \pm 0.1)\times 10^{-11}$  &0.89                        &0.95                        &0.93 & 2.9 \\ \hline
$a_{bs} \neq 0$, $g_{bs}=0$, $ g_{L}^{\mu \mu} \neq  g_{R}^{\mu \mu}$  & $ g'_{bs} g_{\mu\mu}=(-3.2\pm 2.5) \times 10^{-10}$ & $ g'_{bs}g'_{\mu\mu}=(-0.1 \pm 0.1)\times 10^{-11}$  &0.85                        &0.97                        &1.05 & 1.6 \\ \hline
$a_{bs} = 0$, $ g_{L}^{\mu\mu} =  g_{R}^{\mu\mu}$  & $g_{bs}g_{\mu\mu} = (4.4\pm 1.4)\times 10^{-8}$ & $g'_{bs}g_{\mu\mu} =(-1.9 \pm 1.4)\times 10^{-8}$  & 0.86  &0.72 &0.76& 4.6 \\ \hline
$a_{bs} = 0$, $g'_{bs}=0$, $ g_{L}^{\mu\mu} \neq  g_{R}^{\mu\mu}$  & $ g_{bs} g_{\mu\mu}=(3.9\pm 1.8) \times 10^{-8}$ & $ g_{bs}g'_{\mu\mu}=(4.4 \pm 4.2)\times 10^{-11}$  &0.87                        &0.80                        &0.69 & 4.4 \\ \hline
$a_{bs} = 0$, $g_{bs}=0$, $ g_{L}^{\mu\mu} \neq  g_{R}^{\mu\mu}$  & $ g'_{bs} g_{\mu\mu}=(-0.5\pm 5.6) \times 10^{-9}$ & $ g'_{bs}g'_{\mu\mu}=(0.0 \pm 1.5)\times 10^{-11}$  &0.92                        &0.99                        &1.01 & 0.1 \\ \hline
\multicolumn{7}{|c|}{(iii) Light vector with electron coupling  }   \\ \hline
$F(q^2) \equiv 1$, $ g_{L}^{ee} =  g_{R}^{ee}=2.5\times 10^{-4}$ & $g_{bs}g_{ee} = (-0.6 \pm 1.0) \times 10^{-10}$ & $ g_{bs}'g_{ee} = (-0.4 \pm 1.1 ) \times 10^{-10}$ & 0.93                        &0.99                        &0.99  & 0.7\\ \hline
$a_{bs} \neq 0$, $ g_{L}^{ee} =  g_{R}^{ee}$  & $g_{bs}g_{ee} = (-1.9 \pm 0.6)\times 10^{-9}$ & $ g'_{bs}g_{ee}=(-0.8 \pm 0.5)\times 10^{-9}$  & 0.62  &0.92 &0.74& 4.5 \\ \hline
$a_{bs} \neq 0$, $g'_{bs}=0$, $ g_{L}^{ee} \neq  g_{R}^{ee}$  & $ g_{bs}g_{ee} =(-4.4\pm 5.9) \times 10^{-10}$ & $ g_{bs}g'_{ee}=(7.5 \pm 3.3)\times 10^{-10}$  &0.55                        &0.86                        &0.84 & 4.5 \\ \hline
$a_{bs} \neq 0$, $g_{bs}=0$, $ g_{L}^{ee} \neq  g_{R}^{ee}$  & $ g'_{bs}g_{ee} =(3.9\pm 4.2) \times 10^{-10}$ & $ g'_{bs}g'_{ee}=(12.4 \pm 2.6)\times 10^{-10}$  &0.58                        &0.98                        &0.81 & 4.0 \\ \hline
$a_{bs} = 0$, $ g_{L}^{ee} =  g_{R}^{ee}$  & $g_{bs}g_{ee} = (-3.9\pm 1.0)\times 10^{-8}$ & $g'_{bs}g_{ee} =(1.4 \pm 1.0)\times 10^{-8}$  & 0.78  &0.60 &0.75& 4.8 \\ \hline
$a_{bs} = 0$, $g'_{bs}=0$, $ g_{L}^{ee} \neq  g_{R}^{ee}$   & $ g_{bs} g_{ee}=(-3.2\pm 2.3) \times 10^{-8}$ & $ g_{bs}g'_{ee}=(0.4 \pm 1.4)\times 10^{-8}$  &0.83                        &0.70                        &0.67 & 4.6 \\ \hline
$a_{bs} = 0$, $g_{bs}=0$, $ g_{L}^{ee} \neq  g_{R}^{ee}$   & $ g'_{bs} g_{ee}=(4.6\pm 1.5) \times 10^{-8}$ & $ g'_{bs}g'_{ee}=(2.0 \pm 0.3)\times 10^{-8}$  &0.80                        &0.58                        &0.77 & 4.7 \\ \hline
\end{tabular}

}
\end{center}
}
\end{table}

The best fit values of the couplings $g_{bs}^S$ and $g_{bs}^{S'}$ along with 
predictions at the best fit point, for $M_S =25 $ 
MeV and $g_{ee}^S =2.0 \times 10^{-4}$, are 
provided in Table~\ref{tab:scalar}.
%
As a good fit is obtained in this case, we check if these values 
are consistent with the various measured branching ratios in $\bsee$ modes.
If $S$ can decay to $ e^+ e^-$ with a branching ratio $ \sim $1 then the decays $B \to K^{(*)} e^+ e^-$ will be dominated by the two-body decays, $B \to K^{(*)} S$, with $S$ decaying to $ e^+ e^-$.

For the two body $B\to KS$ decay, the branching ratio is
\beq
\mathcal{B}(B\to K S)=\frac{(g_{bs}^S+g_{bs}^{S'})^2|\vec{p}_{K}|(m_B^2-m_K^2)^2f_0^2(m_S^2/m_B^2)\tau_{B}}{32\pi m_b^2m_B^2}\,,
\eeq
where the form factor $f_0(z)$ can be found in Ref.~\cite{Ball:2004ye}.

For the two body $B\to K^{*}S$ decay, the branching ratio is
\beq
\mathcal{B}(B\to K^{*}S)=\frac{(g_{bs}^S-g_{bs}^{S'})^2|\vec{p}_{K^{*}}|^3A_0^2(m_S^2)\tau_{B}}{8\pi m_b^2}\,,
\eeq
where $\tau_{B}$ is the lifetime of $B$ meson, $|\vec{p}_{K^{*}}|=\lambda^{1/2}(m_B^2,m_{K^{*}}^2,m_S^2)/2m_B$, and the form factor $A_0$ is taken 
from Ref.~\cite{Ball:2004rg}.
To bound the NP coupling constants $g_{bs}^S$ and $g_{bs}^{S'}$, we require the $B \to K^{(*)} S$ branching ratio to be less than 1\%. This choice is consistent with uncertainties in the calculation of the $B$ meson width \cite{Lenz_lifetime}. For $M_S$ between $10-200$~MeV, $\mathcal{B}(B^0 \rightarrow K^{*0} e^+ e^-)$ and $\mathcal{B}(B^0 \rightarrow K^{0} e^+ e^-)$ impose the constraints shown in Table~\ref{tab:constraints}.
The best-fit values of the coupling given in Table~\ref{tab:scalar} are in 
contradiction with these constraints. Hence, a light scalar with form factor $F(q^2)\equiv 1$ is ruled out. 

\begin{table}[t]
{
\label{tab:constraints}
\caption{Constraints from  $\mathcal{B}(B^0 \rightarrow K^{0} e^+ e^-)$ and $\mathcal{B}(B^0 \rightarrow K^{*0} e^+ e^-)$. See the text for details. }
\begin{center}
\resizebox{\columnwidth}{!}{%
\renewcommand{\arraystretch}{1.5}
\begin{tabular}{|c|c|c|c|}\hline
  & $\mathcal{B}(B^0 \rightarrow K^{0} e^+ e^-)$ & $\mathcal{B}(B^0 \rightarrow K^{*0} e^+ e^-)$& Combined \\ \hline
$S$, $a_{bs}\neq 0$  & $|g_{bs}^S+g_{bs}^{S'}| \lesssim 9.9 \times 10^{-7}$ & $|g_{bs}^S-g_{bs}^{S'}| \lesssim 9.0 \times 10^{-7}$ & $|g_{bs}^S|\,,\,|g_{bs}^{S'}| \lesssim 9.5 \times 10^{-7}$ \\ \hline
$S$, $a_{bs}=0$   & $|g_{bs}^S+g_{bs}^{S'}| \lesssim 4.4 \times 10^{-2} \left ( \frac{25 ~\rm MeV}{M_{S}} \right )^2$ & $|g_{bs}^S-g_{bs}^{S'}| \lesssim 4.0 \times 10^{-2} \left ( \frac{25 ~\rm MeV}{M_{S}} \right )^2$& $|g_{bs}^S|\,,\,|g_{bs}^{S'}| \lesssim 4.2 \times 10^{-2} \left ( \frac{25 ~\rm MeV}{M_{S}} \right )^2$ \\ \hline
$\Z$, $a_{bs}\neq0$  & 
 $|g_{bs} + g_{bs}'| \lesssim  5.8 \times 10^{-9} \left ( \frac{M_{Z'}}{25 ~\rm MeV} \right )$ & $|g_{bs} - g_{bs}'| \lesssim  5.4 \times 10^{-9} \left ( \frac{M_{Z'}}{25 ~\rm MeV} \right )$  &
$|g_{bs}|,~  |g_{bs}'| \lesssim 5.6 \times 10^{-9} \left ( \frac{M_{Z'}}{25 ~\rm MeV} \right )$ \\ \hline
$\Z$, $a_{bs}=0$   &
 $|g_{bs} + g_{bs}'| \lesssim  2.6 \times 10^{-4} \left ( \frac{25 ~\rm MeV}{M_{Z'}} \right )$ & $ |g_{bs} - g_{bs}'| \lesssim  2.4 \times 10^{-4} \left ( \frac{25~\rm MeV}{M_{Z'}} \right )$ &
$|g_{bs}|,~  |g_{bs}'| \lesssim 2.5 \times 10^{-4} \left ( \frac{25 ~\rm MeV }{M_{Z'}} \right )$ \\ \hline
\end{tabular}
}
\end{center}
}
\end{table}

\subsection{$F(q^2)\neq 1$}
Now we consider a $q^2$-dependent form factor $F(q^2)\neq 1$ which may be loop induced. For momentum transfer $q^2
\ll m_B^2$, $F(q^2)$ can be expanded as~\cite{ZMeV}
\bea
F(q^2) & = & a_{bs} + b_{bs} \frac{q^2}{m_B^2} + \ldots\,,\
\label{FF}
\eea
where $m_B$ is the $B$-meson mass. 
We do not include the $B_s$ mass difference and ${\mathcal{B} ( B_s \to e^+ e^-})$ as constraints since $F(q^2)$ is unknown for $q^2\sim m_B^2$. We assume that $S$ does not couple to neutrinos so that $B\to K\nu\bar{\nu}$~\cite{BKnunubarBaBar,BKnunubarBelle} does not constrain $a_{bs}$.
 Redefining $a_{bs}g_{bs}^{S}$  as $g_{bs}^{S}$, and $a_{bs}g_{bs}^{S'}$ as $g_{bs}^{S'}$, we perform a $\chi^2$-fit 
to the theoretically clean observables $R_K$ and $R_{K^*}$. 
The best fit values of the couplings
and the predictions for $R_K$ and $R_{K^*}$ are shown in Table~\ref{tab:scalar}.
%
Taking into account the constraints on $g_{bs}^S$ and $g_{bs}^{S'}$ from Table~\ref{tab:constraints}
along with the constraints on $g_{ee}$ from the anomalous magnetic moment of the electron, 
we see that the best fit values $\mathcal{O}(10^{-8})$ cannot be achieved in this case.

To avoid the strong constraints from the two-body decays we set $ a_{bs}=0$ in Eq.~\eqref{FF} (thereby also evading the $B\to K\nu\bar{\nu}$ constraint if the mediator couples to neutrinos~\cite{ZMeV}),  and absorbing the factor $b_{bs}$ to redefine $g_{bs}^S$ and $g_{bs}^{S'}$, the matrix element for $\bsee$ is given by
\bea
M_{\bsee}^{S,S'}  = 
 \frac{q^2}{m_B^2}\frac{ g_{ee}^S}
{q^2- M_S^2} \left[ g_{bs}^S(\bar{s} P_L b)  + {g_{bs}^{S'} }
   (\bar{s} P_R b)\right]  (\bar{e}  e) +
 \frac{q^2}{m_B^2} \frac{ g_{ee}^{S'}}
{q^2- M_S^2} \left[ g_{bs}^S(\bar{s} P_L b)  + {g_{bs}^{S'} }
   (\bar{s} P_R b)\right]  (\bar{e} \gamma_5 e)  
    \,.\
\eea

With the form factor $q^2/M_B^2$, requiring $\mathcal{B}(B^0 \rightarrow K^{*0} e^+ e^-)$ and $\mathcal{B}(B^0 \rightarrow K^{0} e^+ e^-)$ to be less than 1\% gives the constraints on $g_{bs}^S$ and $g_{bs}^{S'}$ in Table~\ref{tab:constraints}.
The best-fit values can be found in Table~\ref{tab:scalar}. A reasonable fit is obtained in this case 
with a pull of 4.4.
We see that $\RK$ and $\RKstar$ values in the central $q^2$ bin 
can be reasonably accommodated, while the effect on $\RKstar$ in the low $q^2$ bin is small in this case.
We also evaluated the branching ratios for various $b\to s e^+e^-$ observables; see Table~\ref{fitnewV}.
Our prediction for ${\mathcal{B} ( B\to  K e^+ e^-})_{[1.0-6.0]}$ is somewhat in tension with the experimental result.  Allowing for a 10\% uncertainty in the theoretical prediction~\cite{lat}, the discrepancy is about $2.3\sigma$.
The prediction for the inclusive mode
${\mathcal{B} (B\to  X_s e^+ e^-)}_{[1.0-6.0]}$, which suffers from less hadronic uncertainties, 
is consistent with measurement.

Finally, we considered the case with a pseudoscalar coupling of the electron and find
similar results to that of the scalar coupling.

\begin{table}[t]
\label{fitnew}
\label{fitnewV}
\fontsize{10}{10}\selectfont{
\caption{The experimental results for various $\bsee$ observables, along with predictions for the SM and four new physics cases that fit the $R_K$ and $\RKstar$ data and satisfy the $\mathcal{B}(B\rightarrow K^{(*)} e^+ e^-)$ constraints. The light mediator mass is 25~MeV, $F(q^2) \neq 1$ and $a_{bs} = 0$. }
\begin{center}
\resizebox{\columnwidth}{!}{%
\renewcommand{\arraystretch}{1.5}
\begin{tabular}{|c|c|c||c|c|c|c|c|c|c|}\hline
\multicolumn{3}{|c|}{} &  ${R_K}_{[0.045-1.0]}$  & ${\mathcal{B} ( B\to  K e^+ e^-)}_{[1.0-6.0]}$ & ${\mathcal{B} (B\to  X_s e^+ e^-)}_{[1.0-6.0]}$  & ${B(B^0\to K^{*0}e^+e^-)}_{[0.03^2-1]}$\\ \hline
\multicolumn{3}{|c|}{Experimental results} & -  &$(1.56 \pm 0.18) \times 10^{-7}$~\cite{RKexpt} & $ (1.93 \pm 0.55) \times 10^{-6}$~\cite{Lees:2013nxa} &  $(3.1\pm 0.9)\times 10^{-7}$~\cite{Aaij:2013hha}\\ \hline
\multicolumn{3}{|c|}{Standard model predictions}  &0.98  &$1.69 \times 10^{-7}$ &$1.74 \times 10^{-6}$ & $2.6 \times 10^{-7}$ \\ \hline
\multicolumn{3}{|c|}{\parbox[t]{7cm}{Light scalar \\ $g_{bs}^S g_{ee}^S = 2.7 \times 10^{-8}$,~  $g_{bs}^{S'}g_{ee}^S = -15.5 \times 10^{-8}$}}  &0.93 &$2.5 \times 10^{-7}$ & $2.3 \times 10^{-6}$  &$2.6 \times 10^{-7}$\\ \hline
\multicolumn{3}{|c|}{\parbox[t]{7cm}{Light vector \\ $g_{bs} g_{ee}= -3.9\times 10^{-8}$,~$ g'_{bs}g_{ee}=1.4 \times 10^{-8}$}}  &0.73 &$2.4 \times 10^{-7}$ & $2.6 \times 10^{-6}$  &$2.8 \times 10^{-7}$\\ \hline
\multicolumn{3}{|c|}{\parbox[t]{7cm}{Light vector, $g'_{bs}=0$ \\ $ g_{bs} g_{ee}=-3.2\times 10^{-8}$,~$ g_{bs}g'_{ee}=0.4\times 10^{-8}$}}  &0.66 &$2.7 \times 10^{-7}$ & $2.5 \times 10^{-6}$  &$2.7 \times 10^{-7}$\\ \hline
\multicolumn{3}{|c|}{\parbox[t]{7cm}{Light vector, $g_{bs}=0$ \\ $g'_{bs} g_{ee}=4.6 \times 10^{-8}$,~ $ g'_{bs}g'_{ee}=2.0\times 10^{-8}$}}  &1.04 &$2.4 \times 10^{-7}$ & $2.5 \times 10^{-6}$  &$2.8 \times 10^{-7}$\\ \hline
\end{tabular}

}
\end{center}
}
\end{table}
%


\section{Light $\Z$} 

A $\Z$ with mass less than $2 m_\mu$ was recently proposed in Ref.~\cite{ZMeV} to 
simultaneously explain the measurements of $R_K$ and the anomalous magnetic moment of the muon, with implications
for nonstandard neutrino interactions. Such a $\Z$ may potentially explain $R_{K^*}$ in the
 low $q^2$ bin~\cite{Ghosh:2017ber}.
A $ \Z$  with a mass in the few GeV range was discussed 
recently~\cite{datta_rk,Sala:2017ihs} but the $q^2$ dependence of the WC is not strong 
enough to explain the $R_{K^*}$ at low $q^2$~\cite{datta_rk} .
  Here we focus on an MeV $\Z$.

We assume the flavor-changing $bs\Z$ vertex to have the form,
\beq
F(q^2) \, \bar{s} \gamma^{\mu}\left[ g_{bs} P_L + g_{bs}' P_R \right]  b \, Z^\prime_{\mu} ~.
\label{bsZ}
\eeq
The matrix  elements for $\bsll$ and the mass difference in $B_s$ mixing are
\bea
M_{\bsll}
& = &
\frac{F(q^2)}{q^2- M_{Z'}^2} [\bar s \gamma^{\mu}\left( g_{bs} P_L + g_{bs}' P_R \right) b]
(\bar \ell \gamma^{\mu} (g_{L}^{\ell \ell}P_L + g_{R}^{\ell \ell}P_R) \ell) \nn\\
& - &
 \frac{ F(q^2)}{q^2- M_{Z'}^2} \frac{m_b m_\ell}{M_{\Z}^2}
 (g_{R}^{\ell \ell} - g_{L}^{\ell \ell})
[\bar s \left( g_{bs} P_R + g_{bs}' P_L \right) b]
(\bar \ell \gamma_5 \ell)\,, \nn\\
\Delta M_{{B_s}}^{NP} & = &
\frac{(F(q^2))^2}
  {2 q^2- 2 M_{Z'}^2} \frac{2}{3} f_B^2{m_{B_s}} \left [  \left( g_{bs}^2 +g_{bs}'^2\right) \left( 1- \frac{5}{8} \frac{m_b^2}{M_{\Z}^2} \right )
    - 2 g_{bs}g_{bs}' \left( \frac{5}{6} - \frac{m_b^2}{M_{\Z}^2} \frac{7}{12} \right) \right]\,,    
\label{Zp}
\eea
where we have used Ref.~\cite{soni_2HDM} for $\bs$-$\bsbar$ mixing. Also, we define $g_{\ell\ell}\equiv(g_{L}^{\ell\ell} + g_{R}^{\ell\ell})/2$ and
 $g'_{\ell\ell}\equiv(g_{R}^{\ell\ell} -g_{L}^{\ell\ell})/2$ for convenience.
 

\subsection{$\Z$ with muon coupling}
We begin with the case where the $\Z$ couples to muons and not to the electrons. 

\subsubsection{$F(q^2)\equiv 1$}
We first assume that  $F(q^2)\equiv 1$ and  consider the case 
$ g_{L}^{\mu\mu}=  g_{R}^{\mu\mu} =g_{\mu\mu}$, 
so the leptonic term is a purely vector current.  We perform 
a fit to the $R_K$ and $R_{K^*}$ data, and the new physics 
contribution to the $B_s$ mass difference. We choose $M_{Z'}= 25$~MeV
and fix $g_{\mu\mu}=8.0\times 10^{-4}$, which is the 2$\sigma$ upper bound from 
the anomalous magnetic moment of the muon.
The fit results are 
shown in Table~\ref{tab:Zprime}. We see that the overall improvement 
over the SM is insignificant because $g_{bs}^S$ and $g_{bs}^{S'}$ are suppressed by $B_s$ mixing.

\subsubsection{$F(q^2)\neq 1$}
Now we consider $F(q^2)\neq 1$ and assume an expansion as in Eq.~\eqref{FF}. Keeping only the leading $a_{bs}$ term,  we perform a fit to the observables $R_K$ and $R_{K^*}$ for $M_S= 25$ MeV. We do not employ the new physics contribution to the $B_s$ mass difference as a constraint since $F(q^2)$ is unknown for $q^2\sim m_B^2$. The fit results are shown in Table~\ref{tab:Zprime}. The overall improvement over the SM is poor, with a pull of 2.4. Clearly, a light $\Z$ with pure vector coupling to the muon is unable to explain 
the ${R_K}_{[1.0-6.0]}$, ${R_{K^*}}_{[0.045-1.1]}$ and ${R_{K^*}}_{[1.1-6.0]}$ 
 anomalies simultaneously. However, on removing ${R_{K^*}}_{[0.045-1.1]}$
from the fit, one can easily accommodate  the measured values of ${R_K}_{[1.0-6.0]}$ and ${R_{K^*}}_{[1.1-6.0]}$, and a pull of around 4.0 is obtained.

%

We next consider the case with $a_{bs}\neq 0$ and the $Z'$ also has nonzero axial vector 
coupling with the muons, i.e., $ g_{L}^{\ell \ell} \ne  g_{R}^{\ell \ell}$. To keep the number of new couplings unchanged,
we take either $g_{bs}' = 0$ or $g_{bs}=0$.
%
This case also does not give a good fit to the data; see Table~\ref{tab:Zprime}.

As can be seen from Table~\ref{tab:Zprime}, overall two of the scenarios with $a_{bs}=0$ provide good fits except to the $R_{K^*}$ measurement in the low $q^2$ bin. Morevover, a $\Z$ with purely vector muon coupling is easily compatible with other $\bsll$ 
observables~\cite{datta_rk}.

\subsection{$\Z$ with electron coupling}
We now consider the case where the $\Z$ couples to electrons and not to muons.
\subsubsection{$F(q^2)\equiv 1$}
We first assume that  $F(q^2)\equiv 1$ and we start by considering the case $ g_{L}^{ee}=  g_{R}^{ee} =g_{ee}$ so the leptonic term is a purely vector current. We perform a fit to the $R_K$ and $R_{K^*}$ data, and the new physics contribution to the $B_s$ mass difference. We fix $g_{ee}=2.5\times 10^{-4}$, which is within the 90\% CL upper limit from NA48/2~\cite{Batley:2015lha}. The fit results are shown in Table~\ref{tab:Zprime}. The fit to $R_K$ and $R_{K^*}$ is close to the SM predictions because of  $B_s$ mixing.

\subsubsection{$F(q^2)\neq 1$}
Now we consider $F(q^2)\neq 1$. We fit to the observables $R_K$ and $R_{K^*}$ only since $F(q^2)$ is unknown for $q^2\sim m_B^2$. The best fit results are shown in Table~\ref{tab:Zprime}.
While a good fit to $\RK$ and $\RKstar$ is obtained, we need to check if these couplings
are consistent with other measurements. 
As in the scalar case there is a two-body contribution to
 $\mathcal{B}(B \to K^{(*)} e^+ e^-)$ from $B \to K^{(*)} \Z$ and $\Z$ decaying to $ e^+ e^-$ with a branching ratio $ \sim $1.

The branching ratio for $B \to  K Z'$ is~\cite{Fuyuto:2015gmk,Oh:2009fm},
\begin{equation}
\mathcal{B}(B \to  K Z') = \frac{|g_{bs} + g_{bs}'|^2}{64 \pi} \frac{m_B^2~ \beta^3_{BKZ'}}{M_{Z'}^2 \Gamma_B}  \[f_+^{BK} (M_{Z'}^2) \]^2\,,
\end{equation}
where $\beta_{XYZ} = \lambda^{1/2} (1, M_Y^2/M_X^2,M_Z^2/M_X^2)$ and $f_+^{BK}$ is a form factor.
For $B \to K^* Z'$ the branching ratio is given by,
\begin{equation}
\mathcal{B}(B \to K^* Z') = \frac{\beta_{BK^* Z}}{16 \pi m_B \Gamma_B} \( |H_0|^2 + |H_+|^2  + |H_-|^2\)\,,
\end{equation}
where the helicity amplitudes are defined as,
\begin{equation}
H_0 = (g_{bs} - g_{bs}') \[ -\frac{1}{2} (m_B +M_{K^*}) \xi A_1(M_{Z'}^2) + \frac{M_{K^*} M_{Z'}}{m_B + M_{K^*}}~ \sqrt{\xi^2-1}~ A_2(M_{Z'}^2) \]\,,
\end{equation}
and
\begin{equation}
H_{\pm} =\frac{1}{2} (g_{bs} - g_{bs}') \left [(m_B + M_{K^*}) A_1(M_{Z')}^2 \right] \pm (g_{bs} + g_{bs}') \frac{M_{K^*} M_{Z'}}{m_B + M_{K^*}} \sqrt{\xi^2-1} ~V(M_{Z'}^2)\,.
\end{equation}
$V$, $A_1$ and $A_2$ are form factors~\cite{Ball:2004ye,Ball:2004rg} and 
 $\xi = (m_B^2 - M_{K^*}^2 -M_{Z'}^2)/(2 M_{K^*} M_{Z'})$.

Assuming the decay rate of $B \to  K Z'$ and $B \to  K^* Z'$ to be less than 1\% of the $B^0$ width,
we obtain the constraints shown in Table~\ref{tab:constraints}.
%
%
Since $g_{ee}$ is constrained to be less than $2.5\times 10^{-4}$ at the 90\% CL for $M_{Z'} = 25$ MeV~\cite{Batley:2015lha}, the constraints in 
Table~\ref{tab:constraints} exclude
the best-fit values to explain the $\RK$ and $\RKstar$ measurements in this case. 

We next consider the case when $Z'$ also has nonzero axial vector coupling with the electrons, i.e., $ g_{L}^{ee} \ne  g_{R}^{ee}$. The best-fit results are shown in Table~\ref{tab:Zprime}. 
While a good fit to $R_K$ and $R_{K^*}$ is obtained, the best-fit values do not satisfy the two-body constraints of Table~\ref{tab:constraints} along with the constraint on $g_{ee}$.

Now, to avoid the two-body constraint, like in the scalar case, we set $a_{bs} =0$ in Eq.~\eqref{FF}. 
In this case, assuming $ g_{L}^{ee }=  g_{R}^{ee} =g_{ee}$, i.e., pure vector coupling to 
the electron, and for $M_{Z'} = 25$~MeV,  we fit the product
$g_{ee} g_{bs}$ and $g_{ee} g_{bs}'$ to the $R_K$ and $R_{K^*}$ data. The results are summarized in Table~\ref{tab:Zprime}.
Clearly, at the best fit point the predictions for $R_K$ and $R_{K^*}$ are within the $1 \sigma $ range
of the measurements.
Requiring $\mathcal{B}(B^0 \rightarrow K^{0} e^+ e^-)<1\%$ and $\mathcal{B}(B^0 \rightarrow K^{*0} e^+ e^-)<1\%$, we get the constraints shown in Table~\ref{tab:constraints}.
%
%
The best fit satisfies all constraints on $g_{bs}$, $g_{bs}'$ and $g_{ee}$. From Table~\ref{tab:Zprime}, we see that $\RK$ and $\RKstar$ values in all measured $q^2$ bins 
can be reasonably accommodated. 
We also checked that the predictions for the branching ratios to electron modes are consistent with the various observables; see Table~\ref{fitnewV}.
Our prediction for ${\mathcal{B} ( B\to  K e^+ e^-})_{[1.0-6.0]}$ is somewhat 
higher than the measurement
 and this tension could become significant with a reduction in the theoretical and experimental uncertainties.
The prediction for the inclusive mode
${\mathcal{B} (B\to  X_s e^+ e^-)}_{[1.0-6.0]}$, which suffers from less hadronic uncertainties, 
is consistent with measurement.

Next we consider the case when $Z'$ also has nonzero axial vector 
coupling with the electrons, i.e., $ g_{L}^{ee} \ne  g_{R}^{ee}$. Again, we either set $g_{bs}' =0$ or $g_{bs}=0$. The best-fit values shown in Table~\ref{tab:Zprime} satisfy the constraints on the NP couplings, and the $\RK$ and $\RKstar$ values in all measured $q^2$ bins can be reasonably accommodated. The corresponding branching ratios with electron modes are provided in Table~\ref{fitnewV}. 

\section{Summary} 
In this work we have addressed the recent measurement of 
$\RKstar$ with particular attention to the 
low $q^2$ bin, $ 0.045 \le q^2 \le 1.1 ~{\rm GeV}^2 $.
This measurement has been difficult to explain with 
new physics above the GeV scale. For mediators in the $10-200$~MeV  mass range, we find:
\begin{itemize}
\item[1.] A (pseudo)scalar that only couples to muons cannot explain the $\RK$ and $\RKstar$ measurements as the predicted values are larger than in the SM, in conflict with experiment. An $S$ coupling to only electrons can reproduce the
${R_K}_{[1.0-6.0]}$, ${R_{K*}}_{[0.045-1.1]}$ and ${R_{K*}}_{[1.1-6.0]}$ data,
 but the desired values of the couplings are 
not consistent with the measurements of the branching ratios $\mathcal{B}(B \to K^{(*)} e^+ e^-)$.
A $q^2$-dependent flavor changing $ b-s$  coupling to the scalar can produce compatibility with
$\mathcal{B}(B \to K^{(*)} e^+ e^-)$ and gives a good fit to $\RK$ and $\RKstar$ in the central $q^2$ bin, but the deviation of $\RKstar$ from the SM in the low $q^2$ bin is small.
\item[2.] A $\Z$ with general vector and axial vector couplings to the muon and a $q^2$-dependent $b-s$ coupling 
provides a good fit to the combination of the three
${R_K}$ and ${R_{K*}}$ measurements, but does not fit ${R_{K*}}_{[0.045-1.1]}$ well.
\item[3.] A $\Z$ with general vector and axial vector couplings to the electron
can explain $\RK$ and $\RKstar$ data in all measured bins
but the desired values of the couplings are not consistent with the measurements of $\mathcal{B}(B \to K^{(*)} e^+ e^-)$.
However, a  $q^2$-dependent flavor changing $ b-s$  coupling to the vector is compatible with
$\mathcal{B}(B \to K^{(*)} e^+ e^-)$ and gives good fits to  $\RK$ and $\RKstar$; of the cases we considered, the case with purely vector electron coupling provides the best agreement with the data with a pull of 4.8.
\end{itemize}

\vskip 0.1in
{\it Acknowledgments.} 
We thank W. Altmannshofer, T. Browder, A. Denig, A. Dighe, T. Gershon, D. Ghosh, K. Flood, D. McKeen, G. Miller, L. Piilonen and D. Straub for discussions. A.D. thanks the Institute for the Physics and Mathematics of the Universe for
hospitality and partial support. D.M. thanks the Mainz Institute for
Theoretical Physics (MITP) for its hospitality and partial support during the completion of this work. 
This research was supported by the U.S. NSF under Grant No.
PHY-1414345 and by the U.S. DOE under Grant No. DE-SC0010504.


\end{document}